\documentclass[twocolumn,showpacs,aps,prl,superscriptaddress]{revtex4}

\usepackage{graphicx}
\usepackage{dcolumn}
\usepackage{amsmath}
\usepackage{units}

\input{pubboard/babarsym}

\newcommand{\microns}{\ensuremath{\mu{\rm m}}}
\newcommand{\Btag}{\ensuremath{\B_\mathrm{tag}}\xspace}
\newcommand{\Brec}{\ensuremath{\B_\mathrm{rec}}\xspace}

\newcommand{\Bztoksksks} {\ensuremath{\Bz \to \KS\KS\KS}\xspace}
\newcommand{\cksksks} {\ensuremath{C}\xspace}
\newcommand{\sksksks} {\ensuremath{S}\xspace}

\newcommand{\mmiss}{\ensuremath{m_{\rm miss}}\xspace}
\newcommand{\mb}{\ensuremath{m_{B}}\xspace}
\newcommand{\Bflav} {\ensuremath{B_{\rm flav}}\xspace}
\newcommand{\zrec}{\ensuremath{z_{\CP}}\xspace}
\newcommand{\ztag}{\ensuremath{z_\mathrm{tag}}\xspace}
\newcommand{\ttag}{\ensuremath{t_\mathrm{tag}}\xspace}

\newcommand{\BABARPubYear}    {07}
\newcommand{\BABARPubNumber}  {001}
\newcommand{\SLACPubNumber} {12366}

\def\figurebox#1#2#3{%
    \def\arg{#3}%
    \ifx\arg\empty
    {\hfill\vbox{\hsize#2\hrule\hbox to #2{\vrule\hfill\vbox to #1{\hsize#2\vfill}\vrule}\hrule}\hfill}%
    \else
    {\hfill\epsfbox{#3}\hfill}%
    \fi}

\long\def\inst#1{\par\nobreak\kern 4pt\nobreak
    {\it #1}\par\vskip 10pt plus 3pt minus 3pt}

\begin{document}

\preprint{\babar-PUB-\BABARPubYear/\BABARPubNumber}
\preprint{SLAC-PUB-\SLACPubNumber}

\begin{flushleft}
%  \babar\ Analysis Document \#1644, Version 9\\
   \babar-PUB-\BABARPubYear/\BABARPubNumber\\
   SLAC-PUB-\SLACPubNumber\\
%   hep-ex/\LANLNumber\\[10mm]
\end{flushleft}

\title{
  { \Large \bf \boldmath Measurement of \CP\ Asymmetries in \Bztoksksks Decays }
}
%% author list as of 05-Jan-2007 (582 authors)
%
\author{B.~Aubert}
\author{M.~Bona}
\author{D.~Boutigny}
\author{Y.~Karyotakis}
\author{J.~P.~Lees}
\author{V.~Poireau}
\author{X.~Prudent}
\author{V.~Tisserand}
\author{A.~Zghiche}
\affiliation{Laboratoire de Physique des Particules, IN2P3/CNRS et Universit\'e de Savoie, F-74941 Annecy-Le-Vieux, France }
\author{J.~Garra~Tico}
\author{E.~Grauges}
\affiliation{Universitat de Barcelona, Facultat de Fisica, Departament ECM, E-08028 Barcelona, Spain }
% ...
\collaboration{The \babar\ Collaboration}
\noaffiliation

%\collaboration{The \babar\ Collaboration}

\date{\today}

\begin{abstract}
  We present measurements of the 
  time-dependent \CP-violating 
  asymmetries in \Bztoksksks\ decays based on 384 million $\Y4S\to\BB$
  decays collected with the \babar\ detector at the PEP-II
  asymmetric-energy $B$ Factory at SLAC. 
  We obtain the \CP asymmetry parameters 
  $\cksksks =  0.02 \pm 0.21 \pm 0.05$ and 
  $\sksksks = -0.71 \pm 0.24 \pm 0.04$, where the first
  uncertainties are statistical and the second systematic.
  These results are consistent with standard model expectations.
\end{abstract}


\maketitle

%%%%%%%% Intro paragraph

   In the standard model (SM) of particle physics, the 
   decays \Bztoksksks are dominated by the $b \to s \bar s s$ gluonic
   penguin amplitude.
   A large violation of \CP symmetry is predicted by
   the SM in the proper-time dependence
   of $b \to c\cbar\s$ decays of neutral \B mesons.
   Recent measurements
   of \CP violation in $b \to c\cbar\s$ decays~\cite{Aubert:2006aq} 
   are in good agreement with the SM prediction~\cite{ref:PDG2006}.
   The predicted amplitude of this \CP violation (CPV)
   is $\sin2\beta$, where $\beta = {\rm arg}(-V_{cd}V_{cb}^{\ast}/V_{td}V_{tb}^{\ast})$
   is defined in terms of the elements $V_{ij}$ of the
   Cabibbo-Kobayashi-Maskawa (CKM)~\cite{CKM} quark mixing matrix.
   The SM also predicts that the amplitude of time-dependent
   CPV in $b \to s\qbar\q$ $(\q=d,s)$ decays, defined as $\sin2\beta_{\rm eff}$,
   is approximately equal to $\sin2\beta$.
   Contributions from loops involving non-SM particles
   can give large corrections to the time-dependent CPV amplitudes
   for these decays.
   The theoretical uncertainty in the SM prediction of
   $\sin2\beta_{\rm eff}$ is particularly small, less than 4\%, for the decay
   \Bztoksksks, which is a pure \CP-even eigenstate~\cite{ref:gershon}.
   A violation of $\sin2\beta_{\rm eff} \simeq \sin2\beta$
   would be a clear sign of physics beyond the SM~\cite{Grossman:1996ke}.
   In this paper we present a measurement of the time-dependent \CP-violating
   asymmetries in the decay \Bztoksksks \cite{ref:cc}.

   % --- dataset & babar --- %
   %
   The results presented here are based on $383.6 \pm 4.2$ million $\Y4S\to\BB$
   decays collected with the \babar\ detector at the
   PEP-II asymmetric-energy $\epem$ collider, located at the Stanford Linear Accelerator Center. 
   The \babar\ detector~\cite{ref:babar}
   measures the trajectories of charged particles with
   a five-layer double-sided silicon microstrip
   detector (SVT) and a 40-layer central drift chamber (DCH), both
   operating in a uniform \unit[1.5]{T} magnetic field.
   Charged kaons and pions are identified using measurements of
   particle energy-loss in the SVT and DCH,
   and of the Cherenkov cone angle
   in a detector of internally reflected Cherenkov light.  
   A segmented CsI(Tl) electromagnetic calorimeter (EMC) provides
   photon detection and electron identification.  
   Finally, the instrumented flux return of the magnet allows discrimination
   of muons from pions.

   % --- time dependent CP-asymmetries --- %
   The time-dependent \CP asymmetries are functions
   of the proper-time difference $\deltat\equiv t_{\CP}-\ttag$
   between a fully reconstructed \Bztoksksks{} decay ($B_{\CP}$) and
   the other \B{} meson decay in the event (\Btag), which is partially reconstructed.
   The decay rate $f_+$ $(f_-)$ when the tagging meson is a
   \Bz{} (\Bzb ) is given as
   \begin{eqnarray}
     \label{eqn:td}
     \lefteqn{ f_{\pm}(\deltat) \; \propto \; \frac{e^{-|\deltat|/{\tau_{\Bz}}}}{4{\tau_{\Bz}}} \times }\; \\
     && \left[ \: 1 \; \pm \;
       \: S \sin{( \deltamd\deltat)} \mp C \cos{( \deltamd\deltat)} \: \: \right] \; , \nonumber
   \end{eqnarray} 
   \noindent where $\tau_{\Bz}$ is the \Bz lifetime and \deltamd\ is the
   \Bz--\Bzb mixing frequency. 
   The parameters $C$ and $S$ describe the amount of
   \CP violation in decay and in the interference between decays with and without mixing, respectively.
   Neglecting CKM-suppressed decay amplitudes, we expect
   $\sksksks=-\sin2\beta$ and $\cksksks=0$ in the SM.

   % --- definition of sub-modes --- %
   The data are divided into two subsamples,
   one where all three $\KS$ mesons decay into the
   $\pip\pim$ channel ($B_{\CP(+-) }$) and
   another where one of the $\KS$ mesons decays into the $\ppz$ channel,
   while the other two decay into the $\pip\pim$ channel ($B_{\CP(00) }$).
   % 

   % --- selection of secondaries --- %
   We  form $\piz\to\gamma\gamma$ candidates from pairs
   of photon candidates in the EMC.
   An energy deposit in the EMC is determined to be a photon candidate 
   if no track intersects any of its crystals, 
   it has a minimum energy of 50\mev, and
   it has the expected lateral shower shape in the EMC.
   We reconstruct $\KS \to \ppz$ candidates from \piz pairs
   with an invariant mass in the range $480 < m_{\ppz} < 520$\mevcc. 
   We reconstruct $\KS \to \pipi$ candidates from pairs 
   of oppositely charged tracks, originating from a common vertex, with an invariant mass
   within 12 \mevcc (about 4 standard deviations) of the nominal \KS~ mass~\cite{ref:PDG2006}.  
   We also require the decay vertex to be along the expected flight path
   and the significance of the reconstructed flight distance $\tau_{\KS}/\sigma_{\tau_{\KS}}$
   to be larger than 5.

   % --- kinematic variables --- %
   For each  $B_{\CP(+-)}$ candidate two nearly independent kinematic variables are computed;
   the beam-energy-substituted mass 
   $\mes=\sqrt{(s/2+{\bf p}_i \cdot {\bf p}_B)^2/E_i^2 - {\bf p}^2_B}$, 
   and the energy difference $\DeltaE=E^*_B-\sqrt{s}/2$. 
   Here, $(E_i,{\bf p}_i)\equiv q_{\epem}$ is the
   four-momentum of the initial \epem{} system in the laboratory frame and $\sqrt{s} $
   is the center-of-mass energy, while ${\bf p}_B$ is the
   reconstructed momentum of the \Bz{} candidate 
   in the laboratory frame and $E_B^*$ is its
   energy calculated in the \epem{} rest frame. 
   For each $B_{\CP(00)}$ candidate we use two different kinematic variables.
   They are the reconstructed \Bz mass \mb{} and the missing mass 
   $\mmiss = \sqrt{(q_{\epem} - \tilde{q}_B)^2}$, 
   where  $\tilde{q}_B$ is the four-momentum of the
   $B_{\CP(00)}$ candidate after a mass constraint 
   on the \Bz{} meson has been applied. 
   Due to leakage effects in the EMC, which affect 
   the photon energy measurement and therefore 
   the \piz reconstruction, the shape of the \mb distribution 
   is asymmetric around the mean value.
   This results in this combination of variables being less correlated than 
   \DeltaE and \mes, with better background suppression~\cite{ref:k0spi0prl}.

   % --- selection --- %
   For $B_{\CP}$ signal decays, the \mes{}, $\mmiss$  and \mb\ distributions peak near the 
   \Bz{} mass, while the \DeltaE{} distribution peaks near zero.
   For $B_{\CP(+-)}$ candidates, we require \unit[$5.22<\mes<5.30$]{\gevcc} and
   \unit[$|\DeltaE|<120$]{\mev}.
   For $B_{\CP(00) }$ candidates, we require
   \unit[$5.11<\mmiss<5.31$]{\gevcc} and
   \unit[$|\mb - m_B^{PDG}|<150$]{\mevcc},
   where $m_B^{PDG}$ represents the world-average \Bz~ mass \cite{ref:PDG2006}.
   These selection windows include the signal peak
   and a ``sideband'' region which is used for characterization
   of the background.

   % --- continuum background --- %
   The sample of  $ B_{\CP} $ candidates is dominated by random
   $\KS\KS\KS$ combinations from $\epem\to\qqbar$ $(\q=u,d,s,c)$
   fragmentation (the $q\bar q$ continuum).
   We use topological observables to discriminate
   jet-like $\epem\to\qqbar$ events
   from the more spherical \BB{} events. 
   In the \epem{} rest frame we compute the angle $\theta^*_T$ between the thrust
   axis of the $B_{\CP(+-)}$   ($B_{\CP(00)}$) candidate's decay
   products and that of the remaining particles in the event.  
   We require $|\cos\theta^*_T|<0.90$($0.95$), which reduces
   the number of background events by one order of magnitude.
   We also use the Legendre monomials
   $L_0$ and $L_2$, for the characterization of the event shape~\cite{ref:k0spi0prl}.
   The monomials are combined in a Fisher discriminant ${\cal F}$~\cite{ref:k0spi0prl}
   (ratio $l_2=L_{2}/L_{0}$) for $B_{ \CP (+-) } $ ($B_{ \CP (00) } $) candidates,
   and it is used in the maximum-likelihood fit described below.

   % --- multiple candidates --- %
   The average $B_{\CP}$ candidate multiplicity in the $B_{\CP(00)}$ sample
   is approximately $1.7$, coming from multiple $\KS \to \ppz$ combinations.
   In these events, we select the combination with the smallest $\chi^2=\sum_{i}
   (m_i-m_{\KS})^2/\sigma^2_{m_i}$, where $m_i$ ($m_{\KS}$) is the measured
   (world-average) \KS mass~\cite{ref:PDG2006} and $\sigma_{m_i}$ is its estimated uncertainty.
   We use the same method in the  $B_{\CP(+-)}$ sample, where
   only \unit[$1.4$]{\%} of events have more than one $B_{\CP(+-)}$ candidate.

   % --- charm vetoes --- %
   Since  $\Bz\to\chi_{c0,2}\KS$ decays proceed 
   through a $b \to c\cbar\s$ transition, we remove all $B_{ \CP(+-) } $($B_{ \CP(00) }$) 
   candidates with a $\KS\KS$ mass combination within $3\sigma$ ($2\sigma$) 
   of the $\chi_{c0}$ or $\chi_{c2}$ mass.
   After these vetoes, the total reconstruction efficiency,
   including \KS~ branching fractions,
   is about 6\% (3\%) for $B_{ \CP(+-) }$ ($B_{ \CP(00) }$) candidates,
   assuming a uniform Dalitz distribution.

   % --- B background ---
   The remaining background from \BB events is estimated to be negligible
   for the $B_{ \CP(+-) }$ sample and is absorbed into the $q\bar q$ continuum
   component.
   For the $B_{ \CP(00) }$ sample, we extract the yield of  \BB background events
   simultaneously with the signal and $\qqbar$ event yields.

   % --- tagging --- %
   A multivariate tagging algorithm determines the flavor of the
   \Btag meson and classifies it in one
   of seven mutually exclusive tagging categories~\cite{ref:Sin2betaPRD,Aubert:2006aq}.
   They rely upon the presence of prompt leptons, 
   or one or more charged kaons and pions in the event, and have different purities.
   We measure the performance of this algorithm with a data sample
   (\Bflav) of fully reconstructed $\Bz\to D^{(*)-}
   \pip/\rho^+/a_1^+$ decays. 
   The effective tagging efficiency is $Q\equiv\sum_c \eps^c
   (1-2w^c)^2=0.304\pm 0.003$,
   where $\eps^c$ ($w^c$) is the
   efficiency (mistag probability) for events tagged
   in category $c$.

   % --- vertexing --- %
   We compute the proper-time difference
   $\deltat=\deltaz/\gamma\beta c$ using the known
   boost of the \epem{} system and the measured
   separation between the  $B_{\CP}$  and \Btag{} decay vertices along the
   boost direction ($\deltaz=\zrec-\ztag$)~\cite{ref:Sin2betaPRD}.
   For the $B_{\CP}$ decay, where no
   charged particles are produced at the decay vertex,
   we determine the decay point 
   by constraining the \B production vertex to the
   interaction point (IP) in the plane orthogonal to the beam axis
   using only the $\KS\to\pipi$ trajectories.
   The IP position is determined on a 
   run-by-run basis from two-track events.
   We compute \deltat{} and its uncertainty $\sigma_{\deltat}$ from a geometric fit to the
   $\FourS\to\Bz\Bzb$ system that takes into
   account  this IP constraint and  a Gaussian constraint on the sum 
   of the two $B$ decay times ($t_{\CP}+\ttag$)
   to be equal to $2\:\tau_{\Bz}$ with an
   uncertainty of $\sqrt{2}\; \tau_{\Bz}$~\cite{ref:k0spi0prl,treefitter}.  
   %
          %--- Owen: This fact is included two sentences above "using only the
          %          Ks->pi+pi- trajectories.
          %%%% Due to the large number of degrees of freedom of the geometric fit 
          %%%% used to determine the \Brec decay vertex, we favor the convergence of the fit
          %%%% by neglecting the poor informations which come from the $\KS\to\piz\piz$ decays.
   %
   In order to ensure a well-determined vertex separation
   between \Brec and \Btag,
   we exclude events that have the error on \deltat, 
   determined from the vertex fit, \unit[$\sigma_{\deltat}>2.5$]{ps} and
   events with \unit[$\vert\deltat\vert>20$]{ps}.
   The mean uncertainty in \zrec, a convolution of the uncertainty in the interaction 
   region position and the \ztag resolution,
   is 75{\microns}.
   The mean uncertainty on \ztag\ is about 200{\microns}, which
   dominates the $\deltaz$ uncertainty.
   The resulting $\deltaz$ resolution
   is comparable to that in $\Bz\to\jpsi\KS$ decays~\cite{ref:k0spi0prl}.
   Simulation studies and a $\Bz\to\jpsi\KS$ data control sample
   show that the procedure we use to determine the vertex for
   a $B_{\CP}$ decay provides an unbiased
   estimate of \zrec{}~\cite{ref:k0spi0prl}.

   Most events have at least one \KS\ candidate for which both tracks have at least
   one hit in the inner three SVT layers.
   We have verified on simulation and on data control samples that the parameters of 
   the signal \deltat resolution function 
   for these $B_{\CP}$ signal decays are similar to 
   those obtained from the \Bflav sample~\cite{ref:Sin2betaPRD}.
   When at least one \KS\ has tracks
   with hits in the outer two SVT layers but not in the inner three layers,
   the resolution is nearly two times worse
   and the \deltat{} information is not used. 
   %

   % --- maximum likelihood fit --- %
   We extract the event yields and \CP parameters with an unbinned extended maximum-likelihood fit
   to the kinematic, event shape, and \deltat variables.
   For each of the sub-samples $k=1,2$ ($B_{\CP(+-)}, B_{\CP(00)}$) we use:
   \begin{eqnarray}
     {{\cal L}_k = 
       %%% \frac{
       e^{-(\sum_j^n N_j)}
       %%% }{(N_T)!}  
       \times}  
     {\prod_i^{N_T}{
       }
       \sum_j^n  N_j 
       {\cal P}^i_{j}  ,}  
     \nonumber
   \end{eqnarray}
   \noindent where 
   ${\cal P}_j$ is  the probability density function (PDF) for the $j^{th}$ fit 
   component. 
   $N_j$ is  the event yield of each of the $n$ components:
   $N_S$ signal events, $N_{\qqbar}$ continuum $\qqbar$ events  and,
   for $B_{ \CP(00) } $ only,  $N_{B \bar B}$ 
   $B \bar B$ background events;
   $N_T$ is the total number of events  selected.
   For $B_{ \CP(+-) } $ ($B_{ \CP(00) } $) candidates, the PDF  ${\cal P}_j$ is given by the 
   product  of ${\cal P}_j(\mes){\cal P}_j(\DeltaE){\cal P}_j({\cal F})$
   (${\cal P}_j(\mmiss){\cal P}_j(\mb){\cal P}_j({l_2})$) $\times$ 
   $ {\cal P}_j^c(\deltat,\sigma_{\deltat})\eps^c$, summed over the tagging categories $c$.  
   The product ${\cal L}_1{\cal L}_2$ is maximized to determine the
   common \CP asymmetry parameters \sksksks~ and \cksksks~  
   and the values of $N_j$, which are specific to each sub-sample.
   Along with  \sksksks~ and \cksksks, 
   the fit extracts $\eps^c$ and parameters describing the background.

\begin{figure}[htbp]
     \begin{center}
       \begin{tabular}{cc}
         \includegraphics[width=0.5\linewidth]{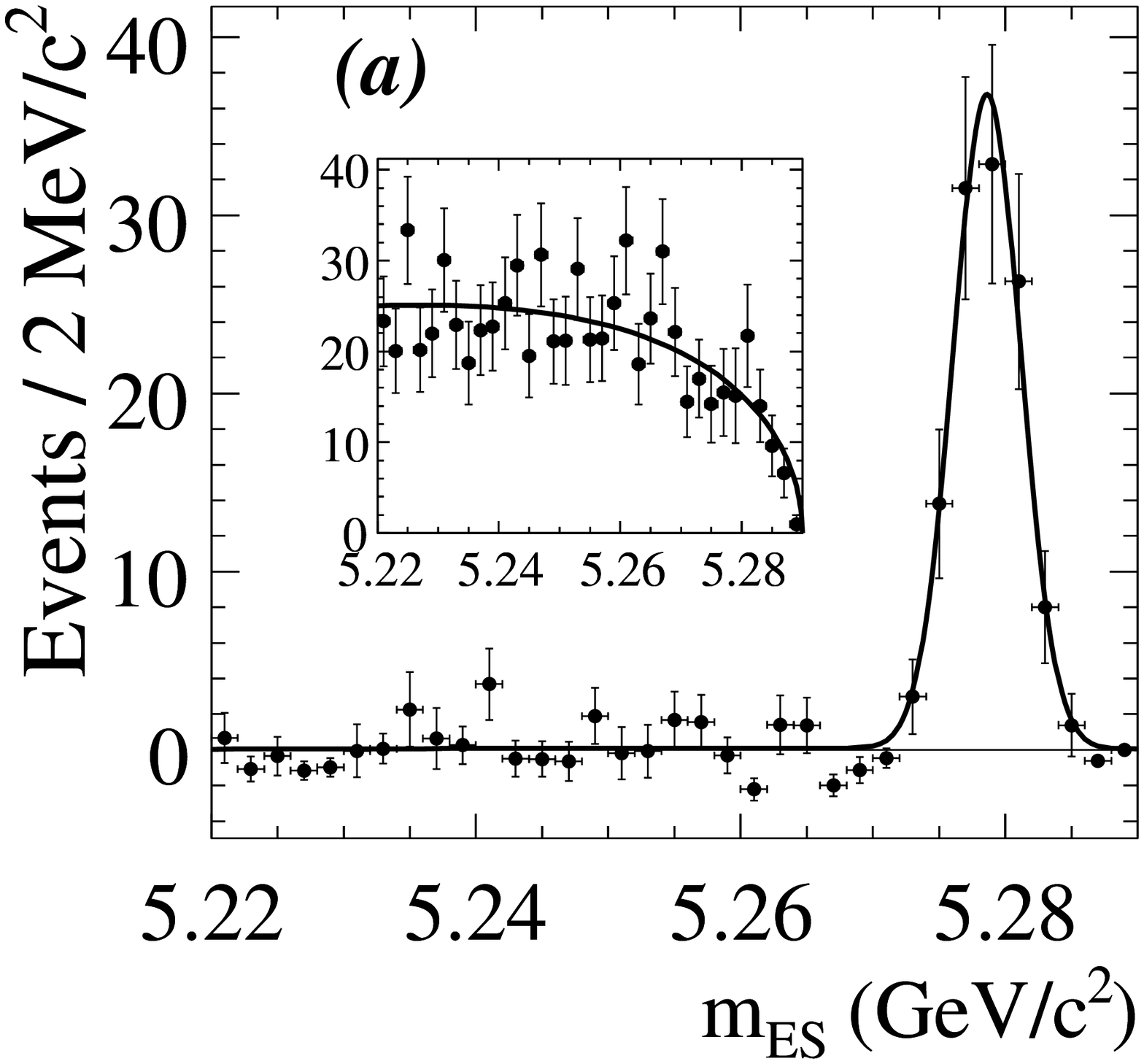} &
         \includegraphics[width=0.5\linewidth]{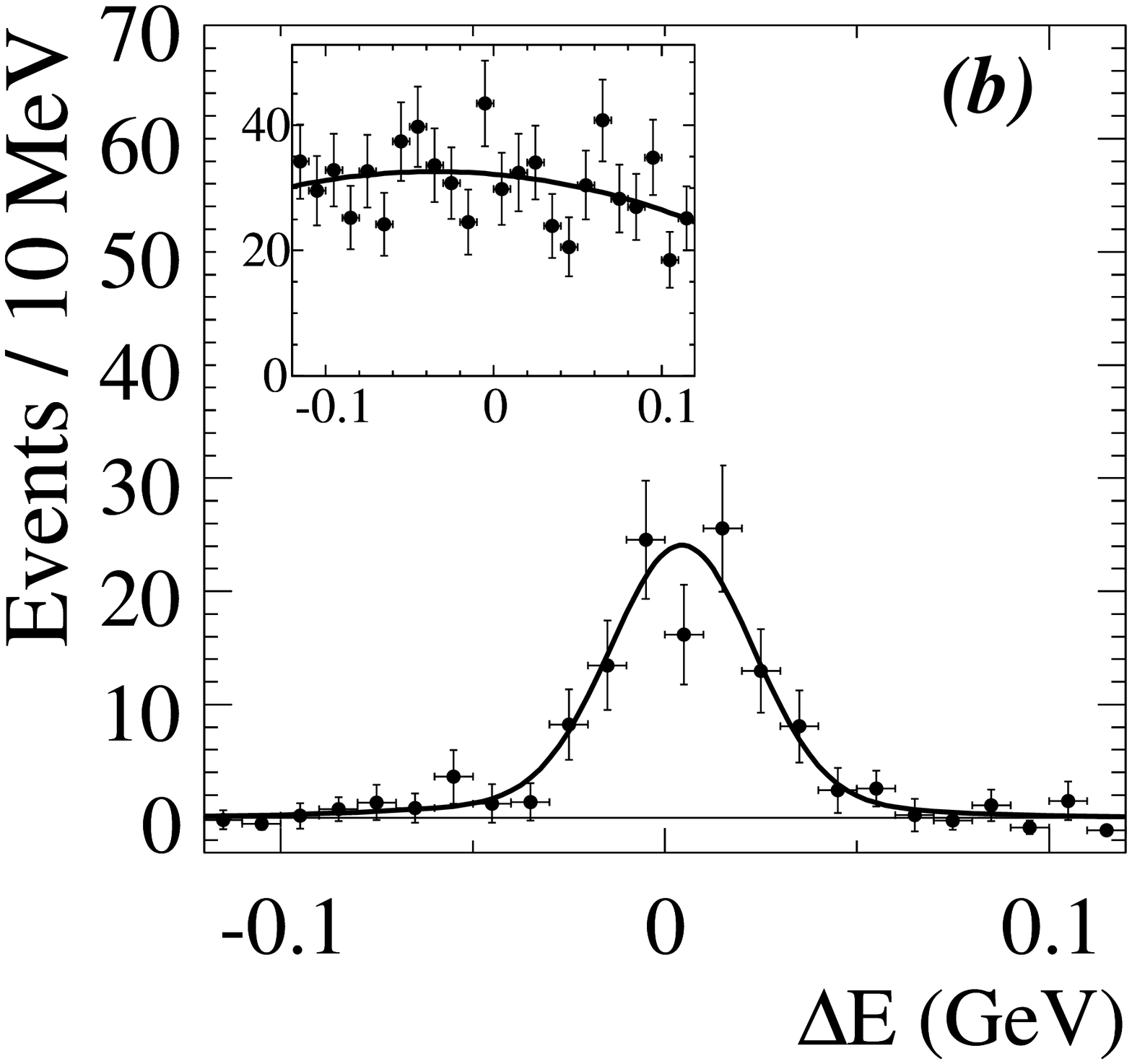} \\
         \includegraphics[width=0.5\linewidth]{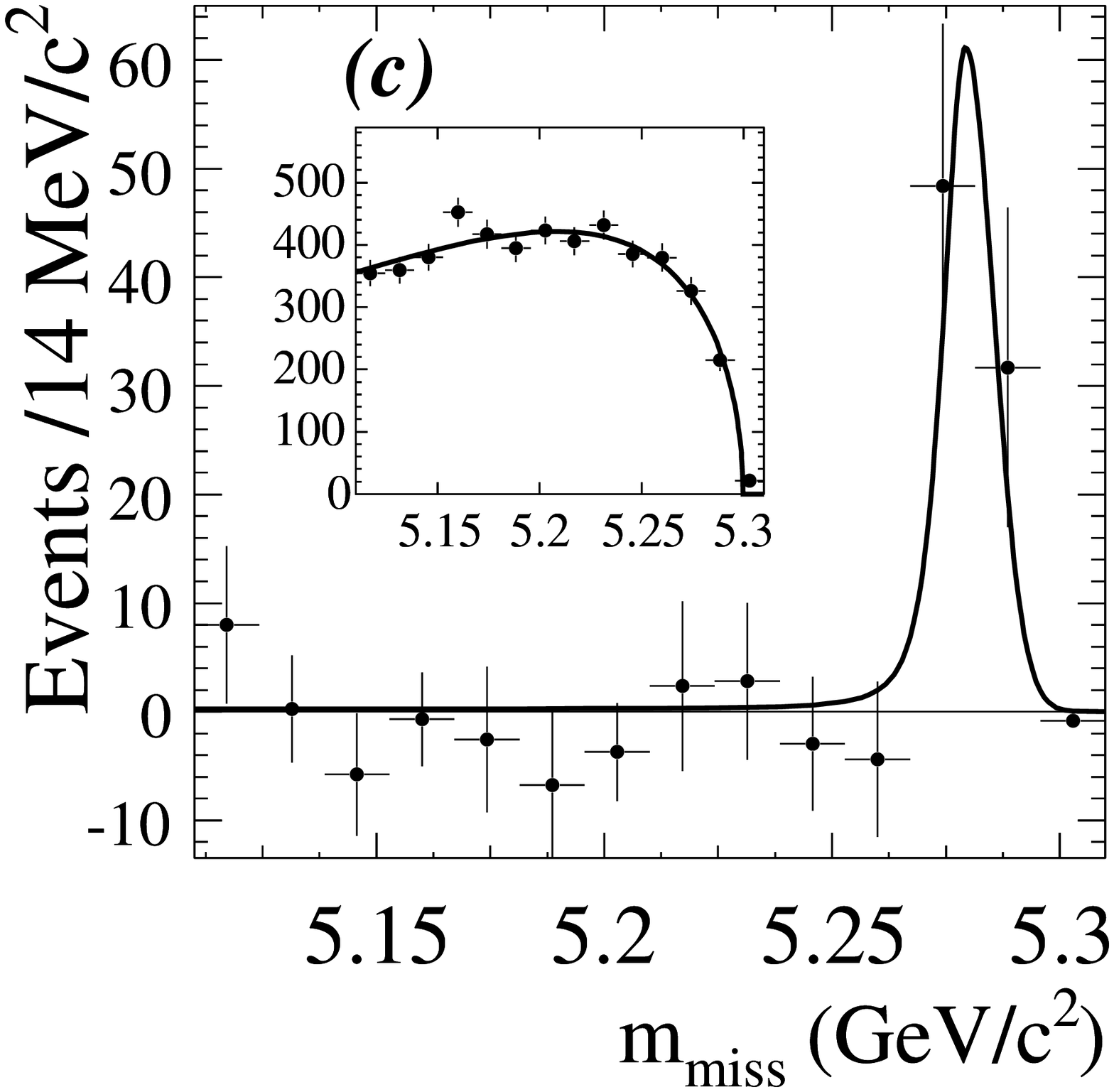} &
         \includegraphics[width=0.5\linewidth]{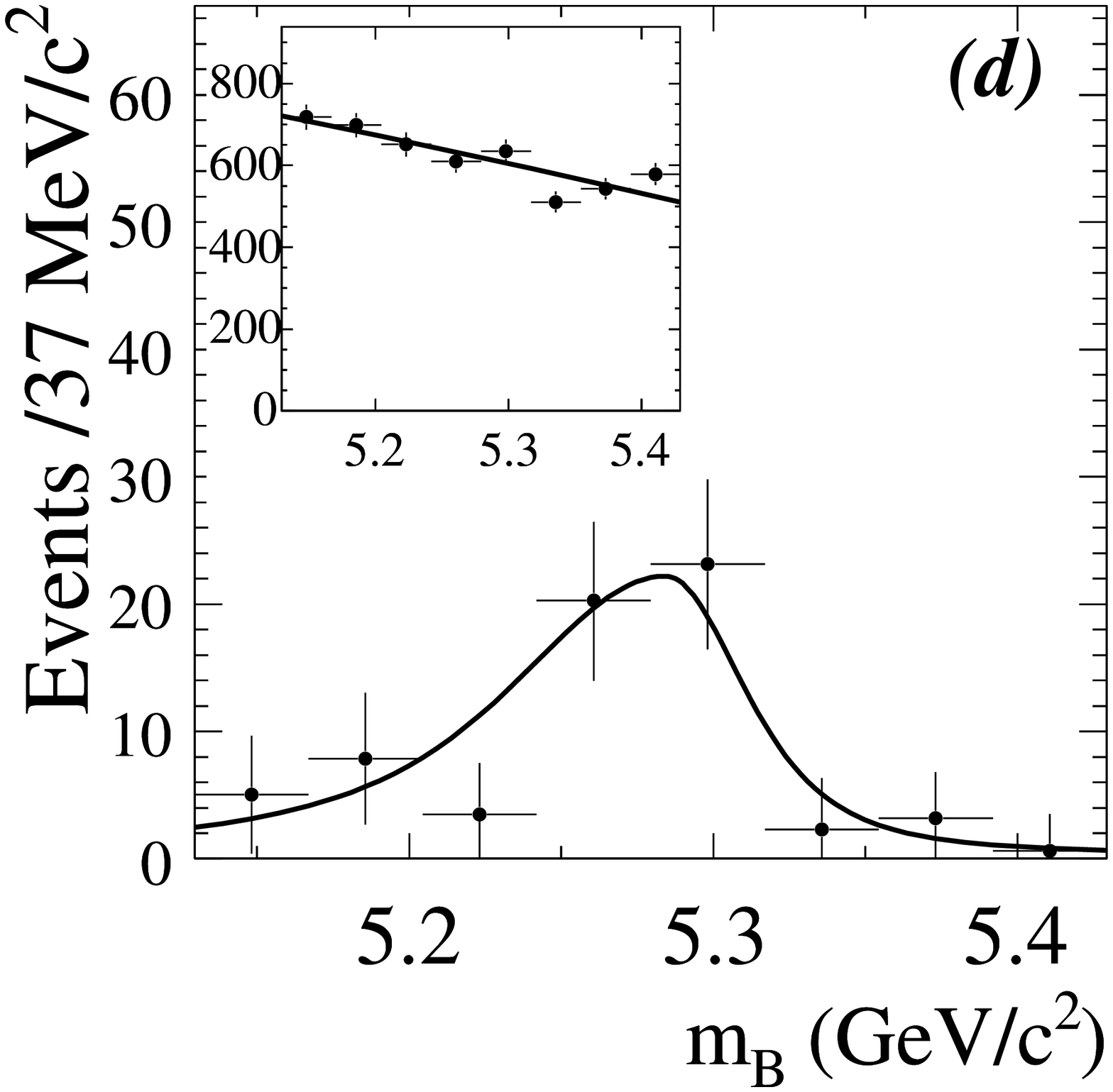} \\
       \end{tabular}
       \caption{ Signal and background distributions of
         (a) $\mes$ and (b) $\DeltaE$ for  $B_{\CP(+-)}$  candidates
         and of (c) $\mmiss$ and (d) \mb \ for $B_{\CP(00)}$ candidates.
         The signal and background distributions have been separated
         using the technique described in~\cite{splot}.
         The curves represent the PDF projections.
         The background distributions are shown in the insets.} 
       \label{fig:selection}
     \end{center}
   \end{figure}

\begin{figure}[!tbp]
     \begin{center}
       \parbox{0.47\textwidth}{\includegraphics[width=0.9\linewidth]{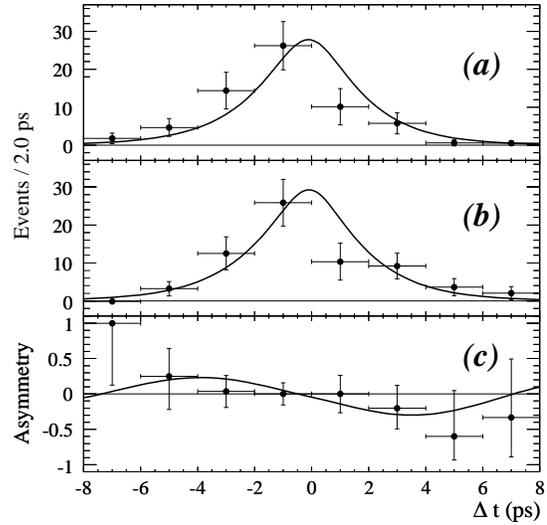}}
     \end{center}
     \caption{
       Distributions of $\deltat$ for events weighted using the
       technique described in \cite{splot} for
       $B_{\rm tag}$ tagged as (a) $\Bz$ or (b) $\Bzb$, and (c) the
       asymmetry ${\cal A}(\deltat)$.  The points are the weighted data
       and the curves are PDF projections. }
     \label{fig:deltat}
   \end{figure}

   % --- fit results --- %
   A fit to 857 $B_{ \CP(+-) }$ and  4992 $B_{ \CP(00) }$ candidates returns the event yields 
   reported in Table~\ref{tab:results}. 
   Figure~\ref{fig:selection} shows the \mes and \DeltaE 
   (\mmiss and \mb) distributions for  signal and background 
   $B_{ \CP(+-) }$ ($B_{ \CP(00) }$) candidates.
   The extracted \CP parameters for the two separate sub-samples 
   and the combined ones are shown in Table~\ref{tab:results}.
   Using a Monte Carlo technique, in which we assume that 
   the measured values for the \CP parameters
   on the combined data sample are the true values, 
   we find that the two sub-samples agree  within  $1.6 \sigma$.
       The statistical significance of the \CP violation is
       evaluated as $\sqrt{2 \cdot \Delta \ln({\cal L}_1{\cal L}_2)}$,
       where $\Delta \ln({\cal L}_1{\cal L}_2)$ is the change in
       the natural log of the combined likelihood for the no \CP-violation
       hypothesis with respect to the maximum value.
   We estimate it to be 2.9 standard deviations.
   \begin{table}[htb]
     \caption{Event yields and \CP\ asymmetry parameters obtained in
       the fit. The errors are statistical only.}
     \centerline{\small
       \begin{tabular}{lcccc}
         \hline\hline
         & \ \ \ \    $B_{\CP(+-)}$  \ \ \ \
         & \ \ \ \    $ B_{\CP(00)}$   \ \ \ \ \ \
         & \ \ \ \    Combined  \ \ \ \   \\ 
         \hline
         $N_S$             & $125 \pm 13$     &   $\phantom{00}64 \pm 12$      & $-$ \\ 
         $N_{\qqbar}$      & $732 \pm 28$     &   $4942 \pm 77$   & $-$ \\ 
         $N_{B \bar B}$   &    $-$      &   $-14 \pm 32$   & $-$ \\ 
         \hline
         \sksksks  \rule[-2.0mm]{0pt}{6mm}  &  $-1.06$ $^{+0.25}_{-0.16}$  &   $0.24 \pm 0.52$  &  $-0.71$ $\pm$ 0.24 \\
         \cksksks  \rule[-2.0mm]{0pt}{6mm}  &  $-0.08$ $^{+0.23}_{-0.22}$  &   $0.23 \pm 0.38$  &   $\phantom{-}0.02$ $\pm$ 0.21 \\ 
         \hline\hline
       \end{tabular}}
     \label{tab:results}
   \end{table}
   Figure~\ref{fig:deltat} shows distributions of
   $\deltat$ for $\Bz$ and $\Bzb$-tagged events, and the asymmetry
   ${\cal
     A}(\deltat) = \left( N_{\Bz} -
     N_{\Bzb}\right) /\left( N_{\Bz} + N_{\Bzb}\right)$. 

   % --- systematic uncertainties --- %
   Systematic uncertainties on the \CP parameters are given 
   in Table \ref{tab:systematics}.
   The systematic errors are evaluated using large samples
   of simulated $B_{\CP}$ decays and the  \Bflav data sample.
   We perform fits to the simulated $B_{\CP}$ signal with parameters
   obtained either from signal or \Bflav\ events to account for 
   possible differences in the \deltat resolution function.
   We use the differences in the resolution function and 
   tagging parameters extracted from these
   samples to vary the signal parameters.
   We account for possible biases due to the vertexing technique by comparing fits 
   to a large simulated sample of IP-constrained (neglecting the \jpsi contribution to 
   the vertex and using the \KS trajectory only) and nominal $\Bz\to\jpsi\KS$ events.
   Several SVT misalignment scenarios are applied to the simulated 
   $B_{\CP}$ events to estimate detector effects.
   We consider variations of 20{\microns} in the 
   direction orthogonal to the beam axis
   for the IP position and resolution and find they have a negligible impact. 
   The systematic error due to correlations
   between the variables used in the fit
   is determined from a fit to a sample of
   randomly selected signal Monte Carlo (MC) events
   added to background events generated from the background PDFs used in the fit.
   The values of the effective \CP parameters for the $B \bar B$ background,
   which are fixed to zero in the nominal fit, are varied over the whole
   physically allowed range.
   The largest deviations in \sksksks and \cksksks resulting from this variation
   are used as systematic uncertainties.
   The world-average values of \deltamd 
   and of the \Bz mean lifetime, $\tau_{\Bz}$, held fixed in the fit, are varied
   by their uncertainties~\cite{ref:PDG2006}.
   We account for the possible interference between the 
   suppressed $\bbar \to \ubar c \dbar$ 
   and the favored $b \to c \ubar d$ amplitudes
   for some \Btag{} decays~\cite{ref:tagint}.
   Finally, we include a systematic uncertainty to account for imperfect knowledge of
   the PDFs used in the fit.  
   Most of this uncertainty is due to MC statistics, the rest to
   differences between data control samples and MC simulation.
   \begin{table}[!htp]
     \caption{Systematic uncertainties on $S$ and $C$.}
     \centerline{\small
       \begin{tabular}{lcccc}
         \hline\hline
         {}              &  \multicolumn{2}{c}{\ \ \ $\sigma (S)$\ \ \ }  
         &  \multicolumn{2}{c}{\ \ \ $\sigma (C)$\ \ \ } \\
         \hline
         Vertex reconstruction         & \multicolumn{2}{c}{0.016}  & \multicolumn{2}{c}{0.003} \\
         Resolution function      & \multicolumn{2}{c}{0.005}  & \multicolumn{2}{c}{0.007} \\
         Flavor tagging          & \multicolumn{2}{c}{0.009}  & \multicolumn{2}{c}{0.015} \\
         SVT alignment and IP position           & \multicolumn{2}{c}{0.016}  & \multicolumn{2}{c}{0.008} \\
         Fit correlation        & \multicolumn{2}{c}{0.004}  & \multicolumn{2}{c}{0.025} \\
         $B \bar B$ \CP, \deltamd and $\tau_{\Bz}$       & \multicolumn{2}{c}{0.008}  & \multicolumn{2}{c}{0.009} \\
         Tag-side interference    & \multicolumn{2}{c}{0.001}  & \multicolumn{2}{c}{0.011} \\
         PDFs        & \multicolumn{2}{c}{0.026}  & \multicolumn{2}{c}{0.031} \\
         \hline
         Total             &\multicolumn{2}{c}{0.037} &\multicolumn{2}{c}{0.046} \\
         \hline\hline
       \end{tabular}}
     \label{tab:systematics}
   \end{table}
   % 

   % --- summary of results --- %
   In summary, we measured the
   \Bztoksksks{} time-dependent \CP asymmetries, 
   \mbox{$\sksksks =
     -0.71 \pm 0.24 \pm 0.04$} and \mbox{$\cksksks =
     0.02 \pm 0.21 \pm 0.05$},
   where the first errors are statistical and the second systematic. 
   The statistical correlation between \sksksks and \cksksks is $-14.1$\%. 
   These results agree well with the SM expectation. 
   %
        %%% The precision achieved 
        %%% by this measurement, which is statistically limited, constraints, but do not exclude, 
        %%% contributions of physics beyond the SM, as the low-energy supersimmetry, described in
        %%% \cite{Grossman:1996ke}.
     %---
      This measurement, which is limited by the small statistics
      of the sample, constrains, but does not exclude
      contributions from physics beyond the SM,
      such as the low-energy supersymmetry~\cite{Grossman:1996ke}.
   These results supersede our previously published \CP asymmetry
   results for \Bztoksksks \cite{prev-babar-3ks-cp}
   and are consistent with the measurements
   performed by the Belle collaboration reported in \cite{Aubert:2006aq}.

   \par
   We are grateful for the excellent luminosity and machine conditions
provided by our \pep2\ colleagues, 
and for the substantial dedicated effort from
the computing organizations that support \babar.
The collaborating institutions wish to thank 
SLAC for its support and kind hospitality. 
This work is supported by
DOE
and NSF (USA),
NSERC (Canada),
CEA and
CNRS-IN2P3
(France),
BMBF and DFG
(Germany),
INFN (Italy),
FOM (The Netherlands),
NFR (Norway),
MIST (Russia),
MEC (Spain), and
STFC (United Kingdom). 
Individuals have received support from the
Marie Curie EIF (European Union) and
the A.~P.~Sloan Foundation.


\begin{thebibliography}{99}
\bibitem{ref:cc} Unless explicitly stated otherwise, charge conjugate decay modes are
     assumed throughout this paper.
\end{thebibliography}
\end{document}